# Electron–Hole Scattering Dichotomy and Anisotropic Warping in Quasi-Two-Dimensional Fermi Surfaces of $UTe_2$

Motoi Kimata[1,*], Jun Ishizuka[2], Freya Husstedt[3,4] Yusei Shimizu[5], Ai Nakamura[5], Dexin Li[6], Yoshiya Homma[6], Atsushi Miyake[6], Yoshinori Haga[1], Hironori Sakai[1], Yoshifumi Tokiwa[1], Shinsaku Kambe[1], Yo Tokunaga[1], Dai Aoki[6], Toni Helm[4,7], and Youichi Yanase[8]

[1]*Advanced Science Research Center, Japan Atomic Energy Agency, Tokai, Ibaraki 319-1195, Japan*

[2]*Faculty of Engineering, Niigata University, Ikarashi, Niigata 950-2181, Japan*

[3]*Institute of Solid State and Materials Physics, Dresden University of Technology, 01062 Dresden, Germany*

[4]*Hochfeld-Magnetlabor Dresden (HLD-EMFL) and Würzburg-Dresden Cluster of Excellence ct.qmat, Helmholtz-Zentrum Dresden-Rossendorf, 01328 Dresden, Germany*

[5]*Institute for Solid State Physics (ISSP), University of Tokyo, Kashiwa, Chiba, 277-8581, Japan*

[6]*Institute for Materials Research, Tohoku University, Oarai, Ibaraki 311-1313, Japan*

[7]*Max Planck Institute for Chemical Physics of Solids, 01187 Dresden, Germany*

[8]*Department of Physics, Graduate School of Science, Kyoto University, Kyoto 606-8502, Japan*

**kimata.motoi@jaea.go.jp*

**Abstract**

We present a combined experimental and theoretical study of the detailed Fermi-surface (FS) geometry of $UTe_2$, a heavy-fermion superconductor that has recently attracted considerable attention as a promising candidate for spin-triplet pairing. Using angle-dependent magnetoresistance oscillations, a bulk- and low-energy-sensitive transport probe for quasi-two-dimensional (Q2D) electronic structures, we directly determine the in-plane FS geometry. We found that the Q2D FS exhibits a rectangular cross-sectional shape with strongly anisotropic warping, originating from the hybridization of two orthogonal quasi-one-dimensional bands. Through a quantitative comparison between experiment and theoretical calculations, we further reveal a large electron-hole scattering dichotomy: the quasiparticle lifetime on the electron FS is substantially shorter than that on the hole FS. This dichotomy is naturally explained by anisotropic, low-dimensional antiferromagnetic fluctuations, which selectively enhance scattering on the electron FS. This suggests a dominant role of the electron pockets for the emergence of superconductivity. Our results clarify a direct relation between FS geometry, magnetic fluctuations, and momentum-dependent quasiparticle lifetimes, and thus providing a crucial basis for the microscopic understanding of pairing mechanism, and impose stringent constraints on the gap symmetry of spin-triplet superconductivity in $UTe_2$.

**Main text**

A detailed understanding of normal-state quasiparticles is crucial for elucidating the pairing mechanism and emergent properties of unconventional superconductivity. In a wide range of correlated superconductors [1], including cuprates [2, 3], ruthenates [4], iron-based systems [5], and heavy-fermion superconductors [6], it is now widely recognized that the orbital character of low-energy electronic states, together with its interplay with magnetic fluctuations, plays a central role. As a crucial first step toward this goal, precise determination of the Fermi-surface (FS) topology is indispensable, as it directly reflects the underlying orbital character and governs the nature of quasiparticle properties.

The recently discovered superconductivity in the heavy-fermion compound $UTe_2$ has therefore attracted considerable attention as a promising platform for spin-triplet pairing [7, 8], which has potential implications for topological superconductivity [9]. Thermodynamic and transport studies of $UTe_2$ have shown that the upper critical fields far exceed the Pauli paramagnetic limit for all crystallographic directions [7], and that multiple superconducting phases emerge under high magnetic fields at both ambient [10, 11, 12, 13, 14] and high-pressure conditions [15, 16, 17]. In addition, microscopic investigations using nuclear magnetic resonance have reported only a small reduction of the spin susceptibility below the superconducting transition [18, 19, 20], together with a rotation of the spin direction (*d*-vector) across different superconducting phases [21, 22]. These observations indicate the presence of internal degrees of freedom within the Cooper pairs, which is a characteristic of spin-triplet superconductivity. Despite extensive efforts to determine its band structure and FS, the electronic structure of $UTe_2$ remains under active debate. In particular, the orbital characters contributing to the low-energy FS, quasiparticle properties, and their relationship to magnetic fluctuations have not yet been experimentally established.

The electronic structure of $UTe_2$ has been investigated by angle-resolved photoemission spectroscopy (ARPES) at $T = 20$ K, however, the reported results are controversial. One ARPES study identified two orthogonal light quasi-one-dimensional (Q1D) bands dispersing along the *a*- and *b*-axes, together with an additional heavy band [23], highlighting the importance of chain-like structures formed by U–U dimers and Te(2) sites running along the *a*- and *b*-directions, respectively (see Fig. 1a). In contrast, another ARPES study emphasized the itinerant character of U 5*f* electrons and the formation of heavy quasiparticles near the Fermi level [24]. These discrepancies may arise from differences in the applied photon energy and/or surface conditions of the measured single crystals.

From the perspective of bulk-sensitive probes operating well below the Kondo temperature, where the heavy-fermion state is fully developed, the FS has been investigated using magnetic quantum oscillations (MQOs) in high magnetic fields; however, the reported results also remain controversial [25, 26, 27, 28]. Several studies revealed two quasi-two-dimensional (Q2D) FS sheets, identified as electron- and hole-like FSs, both characterized by comparably large quasiparticle effective masses of approximately 30–50 $m_0$ ($m_0$: free-electron mass) [25, 26, 27]. In contrast, another study has reported the presence of three-dimensional (3D) FS pockets [28]. Although MQO measurements are a powerful means of probing FS properties, they are primarily sensitive to extremal cross-sectional areas and therefore do not directly provide detailed information on the geometry of the FS cross section in momentum space. Hence, an experimental determination of the detailed FS shape using a bulk- and low-energy-sensitive probe is essential to establish a reliable electronic structure, which forms a crucial basis for a microscopic understanding of spin-triplet superconductivity in this material.

Another important issue to understand the superconducting mechanism is the relationship between magnetic fluctuations and quasiparticle properties in momentum space. In $UTe_2$, inelastic neutron scattering (INS) experiments have revealed incommensurate and low-dimensional antiferromagnetic (AF) fluctuations [29, 30, 31, 32]. However, their influence on the normal-state electronic properties has remained unclear.

Here, we report a bulk- and low-energy-sensitive determination of the FS geometry in $UTe_2$ using angle-dependent magnetoresistance oscillations (AMROs). We identify a Q2D FS with a rectangular cross section in the *ab* plane, that is consistent with hybridization of two Q1D bands. Moreover, a quantitative analysis reveals strongly enhanced quasiparticle scattering selectively on the electron FS, while the hole FS remains largely unaffected. This FS-selective scattering is naturally explained by anisotropic AF fluctuations, establishing a direct link between magnetic correlations and momentum-dependent properties of normal-state quasiparticle in $UTe_2$.

AMROs have proven a powerful tool to probe the electronic structures in a wide range of Q2D systems, including organic conductors [33, 34, 35], cuprates [36, 37, 38, 39, 40], ruthenates [41], and iron-based superconductors [42]. AMRO manifests as oscillations of the *c*-axis magnetoresistance (MR) with respect to the orientation of an applied magnetic field, rather than its magnitude. It is a semi-classical effect that originates from the Boltzmann transport along the lowest-conduction direction in a weakly warped Q2D system. Because AMRO is particularly sensitive to the warping of a Q2D FS, representing the interlayer conduction, systematic field-angle-dependent measurements enable us to extract the in-plane anisotropy of the interlayer electronic dispersion. Therefore, AMRO provides direct experimental approach to the in-plane anisotropy of a Q2D FS and serves as a complementary probe to ARPES and MQO measurements.

The left panel of Fig. 1a shows the crystal structure of $UTe_2$. As indicated by the bonding network, U–U dimers form Q1D chains along the *a* axis, while the Te(2) sites form chains along the *b* axis. These one-dimensional structural networks play a crucial role in shaping the electronic structure of $UTe_2$. The middle and right panels of Fig. 1a illustrate the experimental configuration of the magnetic field direction and a schematic of the measurement setup for the *c*-axis resistivity. To

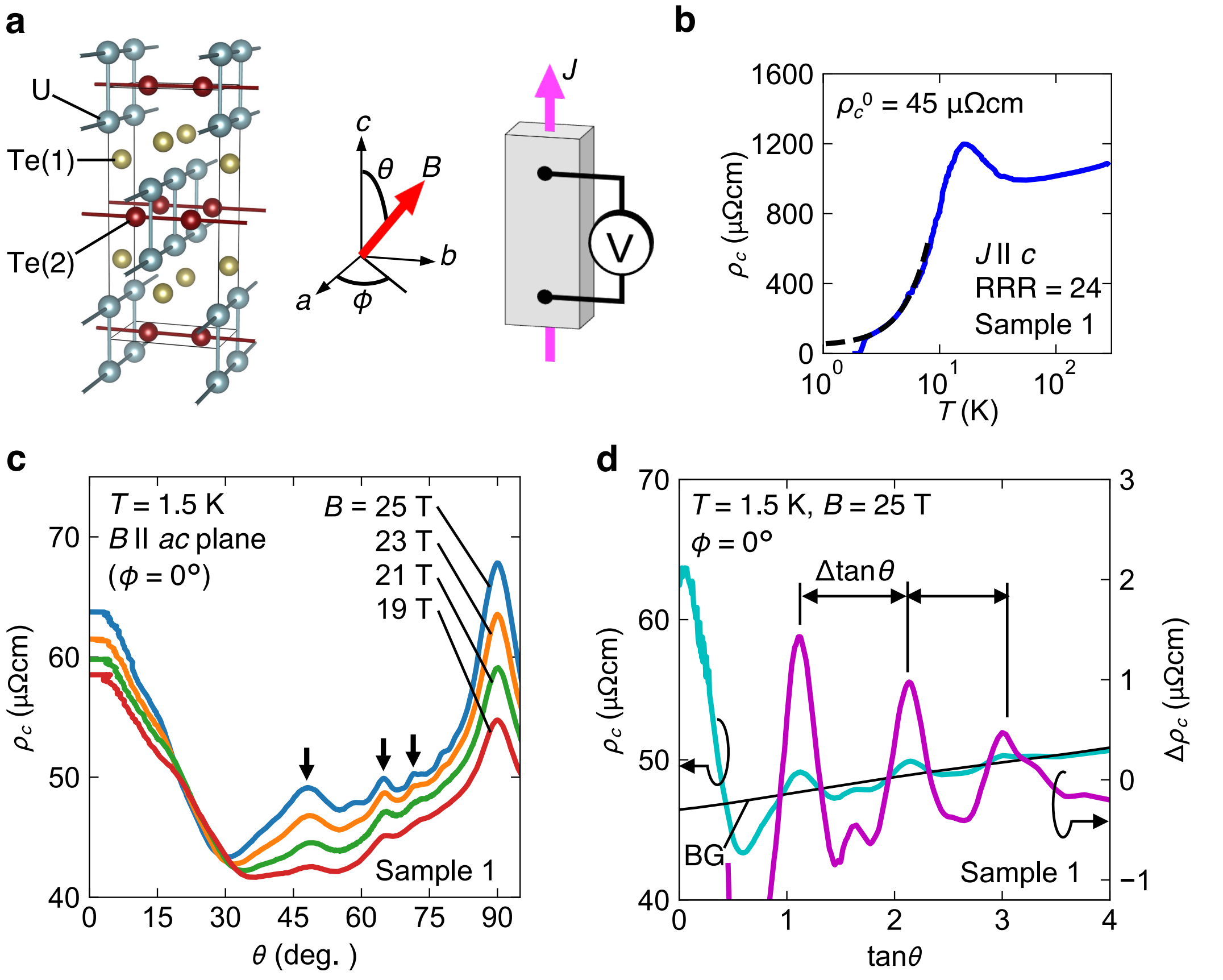

**Fig. 1 | Crystal structure, experimental configuration, temperature dependence of the *c*-axis resistivity, and AMROs in $UTe_2$.**

**a**, (left) Crystal structure of $UTe_2$. As highlighted by the bonding network, chain-like structures formed by U–U dimers and Te(2) sites are running along the *a* and *b* directions, respectively. (center) Schematic of the magnetic-field configuration. The polar angle $\theta$ and the azimuthal angle $\phi$ are defined with respect to the *c* and *a* axes, respectively. (right) Schematic of the experimental setup, where an electrical current *J* is applied along the *c* axis. **b**, Temperature dependence of the *c*-axis resistivity $\rho_c$. The black dashed line represents a fit to $\rho_c = \rho_c^0 + AT^2$, yielding $\rho_c^0$ = 45 μΩ cm and $A$ = 9.4 μΩ cm $K^{-2}$. **c**, $\theta$ dependence of the *c*-axis resistivity $\rho_c$ measured at $T$ = 1.5 K for various magnetic-field strengths with $\phi$ = 0°. The arrows indicate AMRO peaks at $\theta \approx$ 48°, 65°, and 72°, whose positions are independent of the magnetic-field strength. On the other hand, the angular positions of the small kinks observed around $\theta \approx$ 16°–20° are field dependent, which likely reflects the kink induced by the *a*-axis component of the magnetic field associated with the Lifshitz transition observed under an *a*-axis magnetic field [48]. **d**, AMRO data replotted as a function of tan$\theta$. The raw resistivity (cyan, left axis) and background (BG)-subtracted data (magenta, right axis) are shown. In the BG-subtracted data, three pronounced peaks appear, which are nearly periodic in tan$\theta$, further confirming the AMRO origin of the MR oscillations.

observe AMRO, the electric current was applied along the *c* axis. The magnetic field angles were tilted from the *c* axis by a polar angle $\theta$, while the in-plane azimuthal angle $\phi$ was varied with defined from the *c* and *a* axes with polar angle $\theta$ and azimuthal angle $\phi$, respectively. The resistivity was systematically measured as a function of $\theta$ under constant filed and temperature for various values of $\phi$ using a home-built two-axis rotator [43].

Figure 1b presents the temperature dependence of the $c$-axis resistivity. The single crystal investigated in this study was grown by the molten-salt-flux method reported in Ref. 44, and exhibits a superconducting transition temperature of approximately 2.1 K. The residual resistivity ratio (RRR) for electrical current $J \parallel c$ is approximately 24, which is significantly smaller than that for $J \parallel a$ (see Supplementary Information). The residual resistivity ($\rho_c^0$ = 45 μΩcm) is reduced by only a factor of two compared with CVT-grown crystals [45]. This comparison suggests that the $c$-axis resistivity is relatively insensitive to the crystalline quality.

Figure 1c shows the $\theta$ dependence of the $c$-axis resistivity at $\phi$ = 0° for various magnetic-field strength from 19 to 25 T. Clear distinct maxima at $\theta \approx$ 48°, 65°, and 72°, indicated by arrows, are observed. Notably, the peak positions are independent of the field strength. This behavior is characteristic of AMRO, in which the peak positions are solely determined by the geometrical parameters of the Q2D FS, namely the in-plane Fermi wave vector $k_F$ and the warping periodicity along the $c$ axis ($2\pi/c$). The corresponding peak angle of AMRO is known as the Yamaji angles ($\theta_Y$) that satisfy [35]:

$$\tan\theta_Y = (n - \xi)\frac{\pi}{ck_F} \quad (1),$$

where the cross-sectional areas of all cyclotron orbits (perpendicular to the orientation of the external magnetic field) on the Q2D FS become identical in the ideal case. Although real FSs deviate from this idealized condition, the essential effect remains: at the Yamaji angle, the $z$ component of the Fermi velocity is effectively compensated, leading to local maxima in the $c$-axis resistivity. Here, $n$ is an integer ($n$ = 1, 2, 3, … ), indicating that the Yamaji angles appear periodically as a function of tan$\theta$. Also, $\xi$ is the so-called phase factor, which depends on the detailed form of the FS warping [46, 47]. Figure 1d displays the $c$-axis resistivity for $B$ = 25 T and $\phi$ = 0° as a function of tan$\theta$ (left vertical axis). The right vertical axis shows only the oscillatory component after background subtraction. Three clear peaks are visible, nearly periodic in tan$\theta$, which further confirms the AMRO origin of the observed MR oscillations. We also note that the AMROs are reproducibly observed in other crystals (see Supplementary Information).

Next, we show the overall features of the $\phi$ dependence of the AMRO in Fig. 2a. For small azimuthal angles ($\phi < 35°$), clear AMRO peaks are observed, as indicated by arrows. These peaks gradually disappear with increasing $\phi$. At $\phi = 90°$, where the magnetic field is rotated within the *bc* plane, a sharp dip appears around $\theta = 90°$ at $T = 1.5$ K. This dip corresponds to the onset of the high-field superconducting phase, which emerges only when the magnetic field is aligned close to the *b* axis. To avoid the influence of this superconducting transition, we also show the MR curve measured at $T = 2$ K, slightly above the transition temperature of the high-field phase. In this condition, the MR remains in the normal state throughout the full angular range. For this field direction ($\phi = 90°$), no oscillations that can be associated with AMROs are distinguished.

The overall behavior of the MR curves is in good agreement with the theoretical calculation of the *c*-axis resistivity based on the Boltzmann transport equation, as shown in Fig. 2b. The first AMRO peak appears around $\theta \approx 50°$, gradually shifts to larger $\theta$, and eventually vanishes with increasing $\phi$. This behavior is almost consistent with the experimental observations. Although the quantum oscillation experiments and first-principles calculations show two FS sheets [25, 49, 50, 51], which correspond to one hole-like and one electron-like FS sheets, the present calculation (Fig. 2b) assumes that the relaxation time for the hole FS ($\tau_h$) is ten times longer than that of the electron FS ($\tau_e$), i.e., $\tau_h = 1$ ps and $\tau_e = 0.1$ ps. In other words, if the relaxation times for the electron and hole FSs are comparable, a significant AMRO contribution from the electron FS is expected. Theoretical simulations for $\tau_h = \tau_e = 1$ ps indicate that the AMRO signal originating from the electron FS should be most prominent at $\phi = 90°$ [51]. However, no such oscillations are observed experimentally. The good agreement between the theoretical calculations assuming $\tau_h \gg \tau_e$ and the experimental data indicates that the normal-state transport is dominated by the hole FS, whereas the contribution from the electron FS is strongly suppressed. The comparable effective masses of the two FSs revealed by MQO measurements [25] indicate that the cyclotron frequencies for the hole and electron pockets should be similar. This reinforces the conclusion that the relaxation times of hole and electron quasiparticles are substantially different.

The Yamaji angles determined from the experimental data are in good agreement with both the theoretical calculations and MQO results [25] for the hole FS. This provides further confirmation that the observed AMRO originates from the hole pocket. The three pronounced AMRO peaks appear at $\theta \approx 48°$, 65°, and 72° (see Fig.1c). In a Q2D FS, the *c*-axis MR is maximized when the

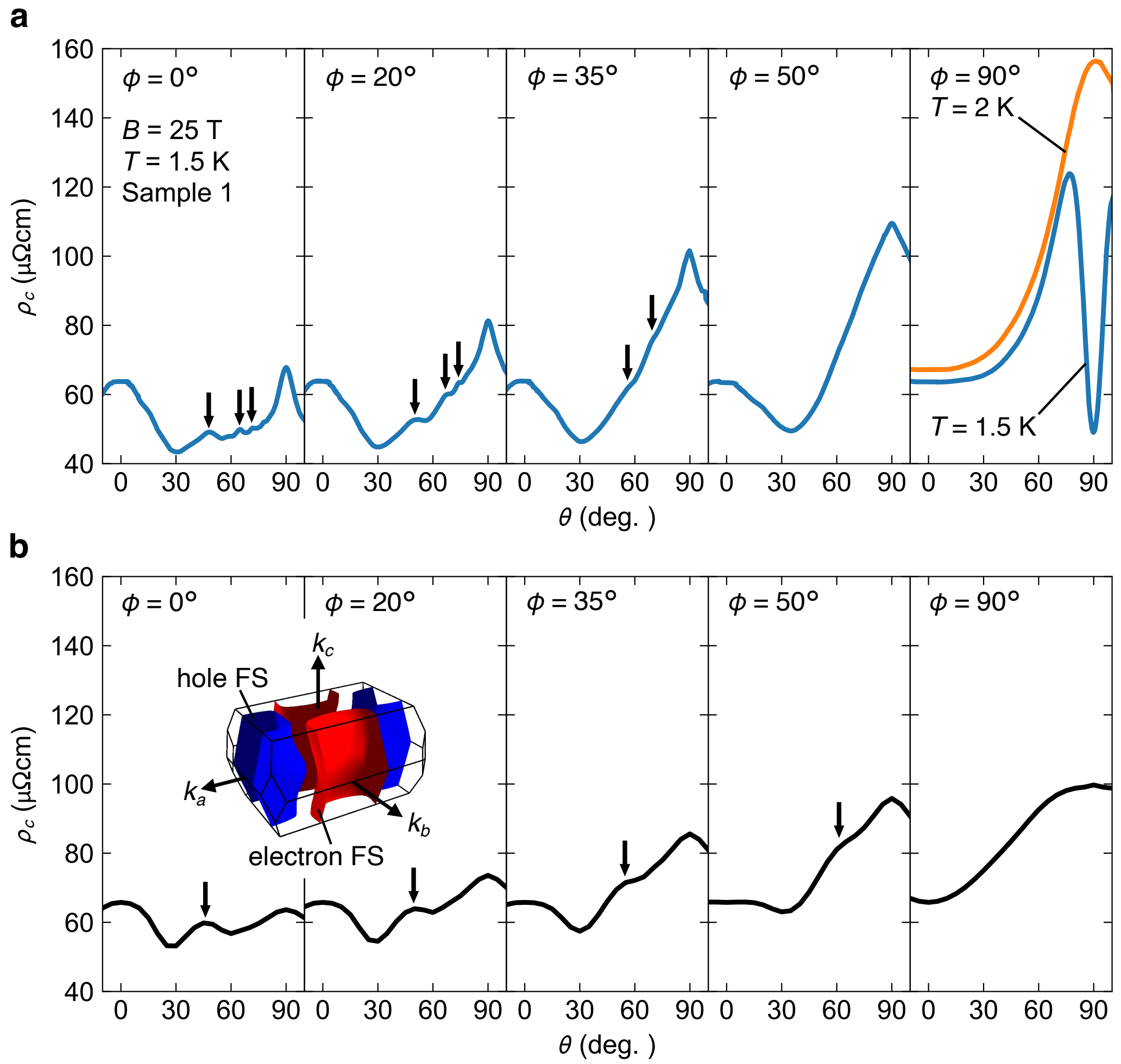

**Fig. 2 | Comparison between experimental and calculated AMROs for various azimuthal feld orientations.**

**a, b,** Experimentally measured (**a**) and theoretically calculated (**b**) AMRO curves for several azimuthal angles $\phi$. (Inset of **b**) Calculated FSs from first-principles calculations. The blue and red sheets represent the hole and electron FSs, respectively. The hole FS exhibits significant $c$-axis warping along the $k_a$ direction, whereas the electron FS shows pronounced $c$-axis warping along $k_b$. In the experimental data at $\phi = 90°$, a sharp dip is observed, originating from the onset of the high-field superconducting transition at $T = 1.5$ K. In both experiment and theory, the AMRO peaks systematically shift to higher polar angles $\theta$ with increasing $\phi$, and gradually broaden and disappear as $\phi$ approaches 90°. The scattering times for the hole and electron FSs in the calculations are set to 1 ps and 0.1 ps, respectively. The magnetic field and temperature used for the calculated AMRO curves are $B$ =12.12 T and $T = 60$ K, chosen to reasonably reproduce the experimental results. Although these parameters differ from the experimental conditions, the overall qualitative behavior is essentially insensitive to the detail of these parameter values.

areas of cyclotron orbits become nearly equal, leading to a merging of the split MQO frequencies. Such merging is theoretically predicted to occur at $\theta \approx 49°$, 67°, and 76° [25], in good correspondence with the experimentally observed Yamaji angles. Notably, the lowest merged angle, which is the most fundamental one, is also observed in the angle-dependent MQO measurements reported previously [25], whereas the higher merging angles are not resolved.

Another key signature reproduced by both experiment and theory is a broad maximum in the $c$-axis MR at $\theta = 0°$. As mentioned above, the $c$-axis MR reaches a maximum when the cross-sectional areas of all cyclotron orbits become nearly equal. Therefore, the appearance of the MR peak at $\theta = 0°$ indicates that this condition is satisfied for this field orientation, suggesting that the corrugation along the $c$ axis is significantly different from a simple cosine dependence. Indeed, such a feature is well established as a characteristic of a staggered-type warping in Q2D systems [53, 54]. It is consistent with first-principles calculations that predicting a staggered $c$-axis warping of the hole FS originating from the body-center type crystal structure of $UTe_2$ [49, 51].

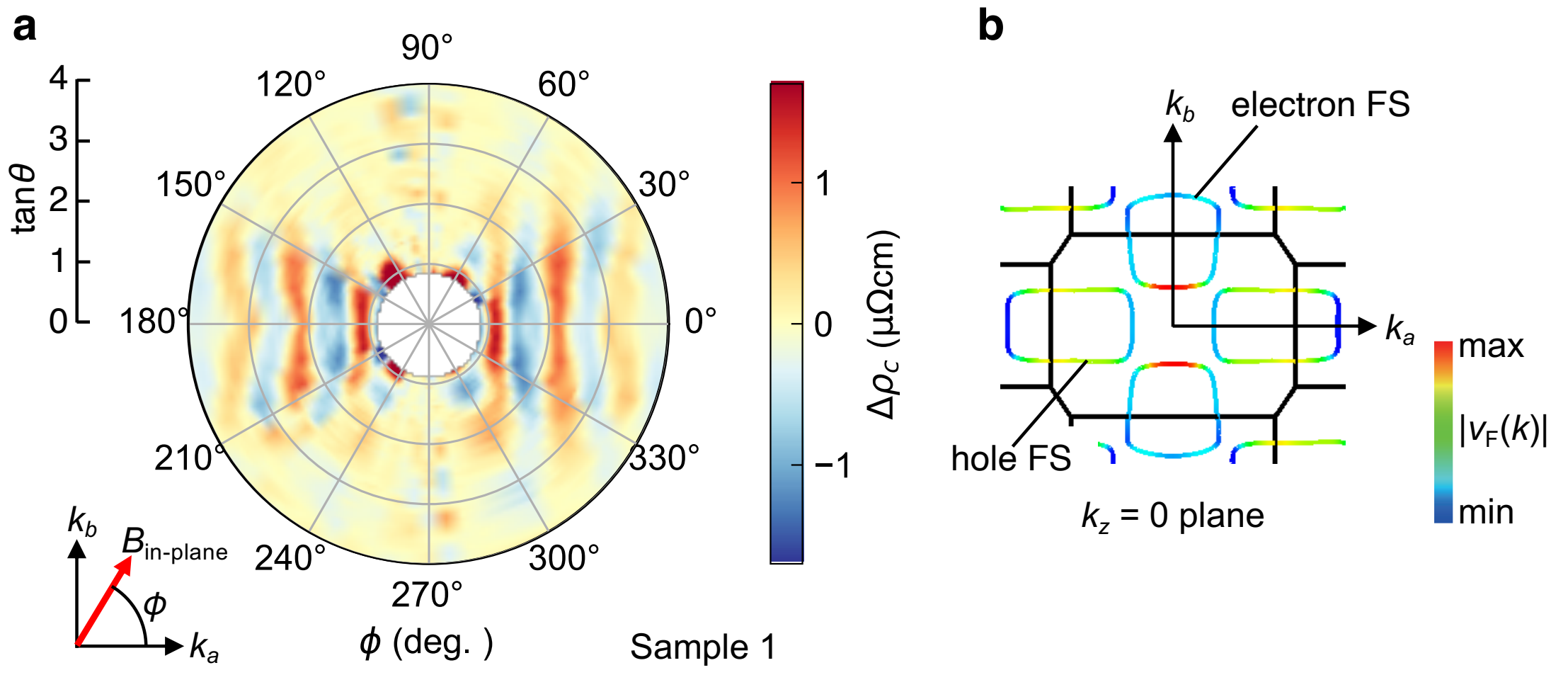

**Fig. 3 | In-plane Fermi-surface geometry of $UTe_2$ determined from AMROs.**
**a**, Two-dimensional color map constructed from experimental AMRO data measured at $B$ = 25 T and $T$ = 1.5 K. A pronounced stripe-like pattern appears only in the angular ranges of approximately −35° (325°) < $\phi$ < 35° and 155° < $\phi$ < 215°. This strong angular selectivity indicates a highly anisotropic $c$-axis warping and a rectangular cross-sectional shape of the Q2D FS in the $ab$ plane. **b**, Cross sections of the hole and electron FSs from first principles calculations at $k_z$ = 0. Both FSs are exhibiting rectangular shapes. The color scale represents the magnitude of the Fermi velocity ($v_F(k)$) at each point on the FSs.

Figure 3a shows a two-dimensional contour plot of the oscillatory component of the $c$-axis MR ($\Delta\rho_c$) as a function of tan$\theta$ with various $\phi$. A periodic contrast associated with AMRO is clearly visible for −35° (325°) < $\phi$ < 35° and 155° < $\phi$ < 215°, consistent with the results in Fig. 2a. A notable feature in this figure is the stripe-like AMRO patterns in the $ab$ plane. This is in sharp contrast to the concentric-ring pattern expected for an in-plane isotropic Q2D FS [55], indicating that the observed Q2D FS is not a simply warped cylinder but has a rectangular cross-section with anisotropic in-plane warping.

Such a rectangular FS shape and anisotropic in-plane warping are consistent with the calculated FSs (see the inset of Fig. 2b and Fig. 3b), and can be understood as the hybridization between two Q1D FS sheets originating from chains of U–U dimers (6*d* orbitals) and Te(2) atoms (5*p* orbitals) running along the *a* and *b* axes, respectively (see Fig. 1a) [49, 50, 51]. First-principles calculations [49, 51] further indicate that the hole FS exhibits strong warping along the $k_a$ direction, while remaining nearly flat along $k_b$. This in-plane anisotropy of warping is fully consistent with our AMRO observations. In contrast, the electron FS is predicted to show the opposite trend, namely significant warping along $k_b$ and much weaker warping along $k_a$. The anisotropic warping of both FSs originates from hybridization between a weak $k_c$ dispersion of a heavy U-5*f* band and two orthogonal Q1D dispersions: the U-6*d* band dispersing predominantly along $k_a$ and the Te(2)-5*p* band dispersing predominantly along $k_b$. A minimal tight-binding model reproduces similarly warped FSs [51]. The strong corrugation of the electron pocket along $k_b$ is therefore expected to give rise to a horizontal stripe-like AMRO pattern. However, no such feature is observed experimentally, as illustrated in Fig. 3a. The absence of this signature in our AMRO measurements further supports the conclusion that the electron pocket makes only a minor contribution to normal-state transport, consistent with a strong-scattering scenario characterized by a short relaxation time for the electron FS, as discussed above.

A quantitative estimation of the *c*-axis warping can also be obtained from the present data. As shown in Fig. 1c, a pronounced AMR peak appears at $\theta$ = 90°, whose width is essentially independent of magnetic field strength. Such a peak in the AMR can be arising from small closed orbits for in-plane field direction, thus it is a hallmark of coherent transport along the *c*-axis [53]. The feature is more clearly resolved in the first-derivative data shown in Fig. 4a. Within the semiclassical framework for a Q2D FS, the peak width around $\theta$ = 90° is given by [56]

$$\delta\theta \approx \frac{2m^* t_c c}{\hbar^2 k_\mathrm{F}} = \frac{t_c}{E_\mathrm{F}} k_\mathrm{F} c \quad (2).$$

Here, $t_c$ and $E_\mathrm{F}$ denote the transfer integral along the *c* axis and the Fermi energy, respectively.

The Fermi energy is given by $E_F = \hbar^2 k_F^2/(2m^*)$, where $m^*$ is the effective mass. The peak width $\delta\theta$ is estimated to be 8 – 10° from the peak-to-peak separation in Fig. 4a, yielding $t_c/E_F \approx$ 0.05 - 0.06 in the present case. Although such a small value of $t_c/E_F$ is consistent with the presence of a Q2D FS, it is relatively large compared with that of typical Q2D conductors, for which $t_c/E_F$ ≈ 0.001 - 0.002 [57, 58], indicating an enhanced $c$-axis warping in the present material. This conclusion is further supported by the anisotropy of resistivity. The upper panel of Fig. 4b displays the temperature dependence of the resistivity for current applied along the $a$ and $c$ axes within the same single crystal, while the lower panel shows the anisotropy ratio $\rho_c/\rho_a$. As shown in this figure, $\rho_c/\rho_a$ remarkably increases below ~20 K, coinciding with the formation of the heavy-fermion state, and reaches a value of approximately 50 at the lowest temperatures. This anisotropy is consistent with the previously reported value in $UTe_2$ [59] and it is substantially smaller than that of typical Q2D conductors [57, 58]. The latter implies that, although the electronic structure retains a Q2D character, it is relatively three-dimensional. This is also consistent with the considerable warping predicted from first principles calculations. Moreover, recent studies of Shubnikov-de Haas quantum oscillations report low-frequency oscillations for magnetic field oriented along the crystallographic $a$ direction, that is perpendicular to the Q2D FS, that are associated with magnetic quantum interference between parts of the electron and hole Q2D FSs. [60, 61, 62]. These findings suggest a significant warping along the $c$ direction confirmed by our observation of AMROs.

We now discuss the possible origin of the substantially different scattering times between the hole and electron FS. The reduced relaxation time of the electron FS may originate from strong AF fluctuations propagating along the $b$ direction, characterized by the wave vector $\mathbf{q} \approx (0, \pi, 0)$, as revealed by INS experiments [29, 30, 31, 32]. This fluctuation predominantly enhances scattering of quasiparticles associated with the corresponding AF wave vector $\mathbf{q}$. As the electron FS has pronounced $c$-axis warping along the $k_b$ direction, such anisotropic magnetic scattering is expected to selectively suppress its contribution to the AMRO. This scenario is consistent with the absence of AMRO features originating from the electron FS in the experimental data.

In $UTe_2$, the crucial role of anisotropic magnetic fluctuations in superconductivity has been established by the observation of a magnetic resonance peak in INS experiments below the superconducting transition temperature [30]. However, their impact on normal-state transport properties has remained largely unexplored. Here, we provide clear evidence for momentum-dependent, and hence FS-dependent, scattering of normal-state quasiparticles driven by magnetic fluctuations. This finding demonstrates that the properties of normal-state quasiparticles are already strongly governed by antiferromagnetic correlations. In particular, the enhanced quasiparticle scattering on the electron pockets suggests their significant contribution to superconductivity, naturally supporting a scenario of magnetically mediated pairing driven by fluctuations at the same fluctuation wave vector **q** [63].

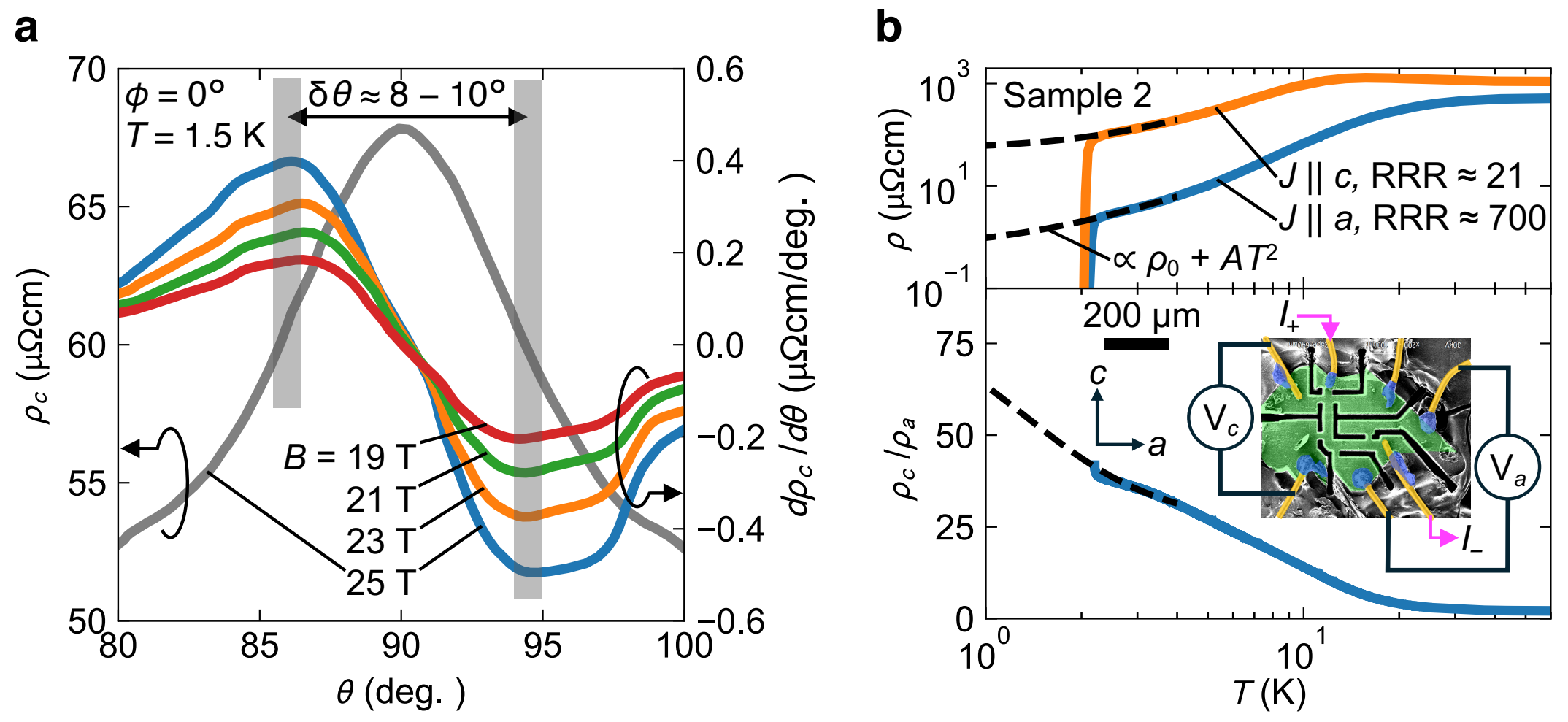

**Fig. 4 | MR peak for in-plane field direction and resistivity anisotropy in $UTe_2$.**
**a**, The $c$-axis resistivity ($\rho_c$) (left axis), and its first derivative with respect to $\theta$ (right axis) measured around $\theta = 90°$ for several magnetic-field strengths. The peak width, estimated from the peak-to-peak separation in the derivative curves, is approximately 8 – 10° and is essentially field independent. **b**, (Upper panel) Temperature dependence of the resistivity along the $a$ and $c$ axes, measured using a focused-ion-beam-fabricated microstructure shown in the inset of the lower panel. A large RRR is observed for $J \parallel a$, whereas that for $J \parallel c$ is significantly smaller. The dashed line indicates an extrapolation to the low-temperature limit based on the $\rho_0 + AT^2$ dependence with $\rho_0^a$ = 0.63 μΩcm, $A$ = 0.35 μΩcm $K^{-2}$ for $J \parallel a$, and $\rho_0^c$ = 53 μΩcm, $A$ = 8.9 μΩcm $K^{-2}$ for $J \parallel c$, respectively. The obtained $\rho_0^c$ and $A$ for $J \parallel c$ are reasonably consistent with those in Fig. 1b. (Lower panel) Temperature dependence of the resistivity anisotropy ratio ($\rho_c/\rho_a$) which increases rapidly below ~20 K.

In summary, we have performed a combined study of AMRO measurements and theoretical simulations based on first-principles-derived FSs. The rectangular in-plane shape of the Q2D FS, directly revealed by AMRO, demonstrates that hybridization between two orthogonal Q1D electronic states deeply persists in the heavy-fermion regime, well below the Kondo temperature. Through a quantitative comparison with theoretical calculations, we further reveal a pronounced and highly selective scattering dichotomy between the electron and hole FSs, with a substantially shorter lifetime on the electron FS. This enhanced scattering on the electron FS is consistently explained by anisotropic, low-dimensional AF fluctuations observed in INS, and suggest a dominant role of the electron pockets in superconductivity. Our findings directly demonstrate a strong link between FS geometry (orbital character) and momentum-dependent quasiparticle scattering driven by magnetic fluctuations, providing an essential foundation for understanding the spin-triplet pairing mechanism, and imposing stringent constraints on the gap symmetry in $UTe_2$.

## Methods

### Single-crystal growth and high-field MR measurements

Single crystals were grown using a molten-salt flux method. The crystallographic orientations were determined by X-ray Laue diffraction. Samples were cut and polished into plate-like geometries for transport measurements with current applied parallel to the c axis. High-field MR measurements, including angle-dependent magnetoresistance oscillations (AMROs), were performed using a standard four-probe low-frequency ac technique at High Field Laboratory for Superconducting Materials (HFLSM), Institute for Materials Research, Tohoku University. Electrical contacts were formed by spot welding Au wires onto the samples, ensuring low and stable contact resistance. AMRO measurements were carried out at temperatures down to $T = 1.5$ K and in magnetic fields up to 25 T. The magnetic-field orientation was controlled using a home-built two-axis rotator. The angular step of the $\theta$ rotation measurements was approximately 0.2°.

### Theoretical calculations

We conducted first-principles calculations for non-magnetic states on $UTe_2$. We adopted the full-potential linearized augmented plane wave plus local orbitals method within the generalized gradient approximation in the WIEN2k package [64, 65]. We set the muffin-tin radius $R_{\mathrm{MT}}$ of 2.50 a.u. and the maximum reciprocal lattice vector $K_{\max}$ as $R_{\mathrm{MT}}K_{\max} = 11.0$. For the GGA+U calculation, we set the Coulomb interaction $U = 2.0$ eV and Hund's coupling $J = 0$ eV. To calculate the conductivity and resistivity under the magnetic field, we constructed a 12-band Wannier model using Wannier90 [66]. The calculated FSs of the Wannier model are shown in Fig. 2b, which reproduce the FSs of the GGA+U calculation. The conductivity and resistivity tensor were calculated using the Boltzmann equation with the semiclassical and relaxation time approximation as implemented in the WannierTools package [67, 68]. As discussed in the main text, it is important to use the band-dependent relaxation time and assume $\tau_h > \tau_e$ to obtain the AMROs that are consistent with the experimental results. Here, we set $\tau_h$ = 1 ps and $\tau_e$ = 0.1 ps.

**Acknowledgments**

We thank K. Machida, H. Harima, M. Shimizu, I. Sheikin, W. Knafo, for fruitful discussions. This work was partly supported by JSPS and MEXT KAKENHI with Projects No. JP25H00599, No. JP25K00955, No. JP24K21522, No. JP23H04868, No. JP23KK0052, No. JP22H00109, No. JP22H04933, No. JP22H01181, No. JP22H04933, No. JP23K17353, No. JP23K22452, No. JP24K21530, No. JP24H00007, No. JP25H01249, No. JP25K07223, No. JP24K00590. Moreover, the work was performed under the GIMRT Program of the Institute for Materials Research, Tohoku University (Proposals No. 202503-HMKPA-0065, No. 202312-HMKPA-0085). T.H. and F. H. acknowledge support from the Deutsche Forschungsgemeinschaft (DFG) Grant No. HE 8556/3-1.

**Author contributions**

M.K. planned the project and designed the experiments. M.K., F. H., and T.H. performed high-magnetic-field experiments and data analysis. J. I. and Y.Y. performed theoretical calculations. A.N. and D.A. grew the single crystals. Y.S., D.L., Y. Homma, and A.M. performed the fundamental characterizations. M.K. and J.I. wrote the manuscript with feedback from T.H., D.A., A.M., Y. Haga, H.S., Y. Tokiwa, S.K., Y. Tokunaga, and Y.Y. All authors discussed the results and contributed to the manuscript.